\def\beg{\begin{equation}}
\def\eeq{\end{equation}}
\documentstyle[12pt]{article}
\begin{document}
\begin{center}
{\Large{\bf The alternative to the incompressible fractional charge in quantum Hall effect: Comments on Laughlin and Schrieffer's papers.}}
\vskip0.35cm
{\bf Keshav N. Shrivastava}
\vskip0.25cm
{\it School of Physics, University of Hyderabad,\\
Hyderabad  500046, India}
\end{center}

Laughlin has found ``exactly" the wave function which is ascribed
to an excitation of fractional charge, such as e/3. We find that
the exactness of the wave function is not destroyed by changing 
the charge to some other quantity, such as the magnetic field. 
Thus e/3 and H can be replaced by e and H/3. Therefore, the wave function need not belong to a quasiparticle of charge e/3.
\vfill
Corresponding author: keshav@mailaps.org\\
Fax: +91-40-3010145.Phone: 3010811.
\newpage
\baselineskip22pt
\noindent {\bf 1.~ Introduction}

     An effort is made to see if there is an alternative to the interpretation of the fractional charge in the quantum Hall effect.
Laughlin has written the wave function for the quasiparticle of fractional charge. We examine to see if the wave function can be interpreted to describe integer charge and the blame of the 
fraction can be thrown on some other quantity. Laughlin$^1$ has 
proposed a variational ground state which is a new state of matter,
a quantum fluid, the elementary excitations of which are 
fractionally charged. The correctness of the wave functions is 
verified by direct numerical diagonalization of the many-body Hamiltonian. We wish to examine the alternative interpretation of
the wave function without destroying the exactness of the
calculation. In particular, we examine whether some other 
variable can take the blame of the fraction instead of the charge.
If we bring a flux quantum near a solenoid, according to Laughlin, charge is affected whereas according to the correct answer, the 
field is affected. The charge density should automatically give 
the correct electric and magnetic vectors so that the magnetic 
field should be automatically correct. There are two variables, 
the charge and the field in the flux quantization. Therefore,
Laughlin's throwing the blame only on the charge is not correct 
and the automatic correction to the electric and magnetic vectors
does not occur.

     We find that several quantities arise as a factor of the 
charge so that the effect of the fraction need not be thrown on
the charge.  Since the charge multiplies the magnetic field, the 
factor arising in the field can also be read as a factor of the 
charge, then there is no trouble in defining a fractional charge. However, if charge is corrected without correcting the magnetic 
field, apparently there is no trouble, except that the 
choice of the variale is not unique.

     In this comment, we explain the various choices available
without disturbing the exactness and point out that field is 
better than the charge.

\noindent{\bf 2.~~Theory}

      Let us consider a two-dimensional electron gas in the $x-y$
plane subjected to a magnetic field along z direction. The eigen
states of the single-body Hamiltonian can be written as, $|m,n>$. The cyclotron frequency is, $\hbar\omega_c$=$\hbar(eH_o/mc)$. The 
magnetic length is $a_o$=($\hbar/m\omega_c$)$^{1/2}$ which upon substituting the cyclotron frequency becomes,
\beg
a_o = ({\hbar c\over eH_o})^{1/2}.
\eeq
The energy levels of the type of a harmonic oscillator are produced 
when the applied field is along the z direction,
\beg
{\it H}|m,n> = ( n + {1\over 2})|m,n>.
\eeq
The wave function of the lowest Landau level is written as a 
function of $z=x+iy$ with $|m>$ as an eigen state of angular 
momentum with eigen value $m$. The many-body Hamiltonian consists 
of, the kinetic energy with the vector potential included in the 
momentum, a potential due to the positively charged nuclei, V(z$_j$), 
and the Coulomb repulsion between electrons. The wave functions 
composed of states in the lowest Landau level which describe the 
angular momentum $m$ about the center of mass are of the form,
\beg
\psi = (z_1-z_2)^m(z_1+z_2)^nexp[(-{1\over 4})(|z_1|^2+|z_2|^2)].
\eeq
Laughlin generalized this observation to $N$ particles by writing product of Jastrow functions,
\beg
\psi = \{\Pi_{j<k}f(z_j-z_k)\}exp(-{1\over 4}\Sigma_l|z_l|^2)
\eeq
which minimizes the energy with respect to $f$. If $\psi$ is antisymmetric $f(z)$ must be an odd function, $f(z)$ = z$^m$ with
 m = {\it odd}. To determine which $m$ minimizes the energy, 
we write,
\beg
|\psi_m|^2 = exp(-\beta\phi)
\eeq
where $\beta$=1/m and $\phi$ is the classical potential energy 
which describes a system of $N$ identical particles of charge 
$Q$=m with the neutralizing back ground charge density,
$\sigma$=(2$\pi$ a$_o^2$)$^{-1}$ per unit area. This is the 
classical one-component plasma (OCP). The solution of which is 
well known. For $\Gamma$=2$\beta$Q$^2$=2m $>$ 140, the OCP is a hexagonal crystal and fluid otherwise. Laughlin's wave function describes
a fluid of density,
\beg
\sigma_m={1\over m(2\pi a_o^2)}
\eeq
which minimizes the energy. The charge density generated by (6)
should be equal to that of the background charge so that there 
is overall charge neutrality.

     If we define a new value of $a_o$, then $ma_o^2$ is replaced 
by $a_{new}^2$. Then the effect of $m$ is not on the charge but it 
is in the distance. That means that the effect of $m$ can occur on, (i)h, (ii)c,(iii) e or (iv) H$_o$. Thus there are at least four candidates to absorb the effect of $m$. Let us eliminate the 
Planck's constant and the velocity of light. Then it is possible 
to affect either $e$ or $H_o$. It will surely be interesting if 
the effect of $m$ is thrown on the velocity of light. Then,
\beg
a_{new}^2={\hbar (mc) \over eH_o}
\eeq
where for $m$=3, the quasiparticles will travel with the velocity of 
3$c$ which is faster than light. In the expression (1), the velocity 
of light $c$ is the value in vacuum so that the particles with 
velocity 3$c$ will be travelling faster than light which will be 
{\it noncausal or causality violating}. Another interpretation is obtained by throwing the value of $m$ on the magnetic field,
\beg
a_{new}^2= {\hbar c\over e(H_o/m)}.
\eeq
Therefore, the number $m$ of Laughlin can be absorbed as a factor 
of magnetic field such as,
\beg
 gH_o=H_o/m.
\eeq
It is perfectly allowed to devide the magnetic field by $m$ to 
define a new field. {\bf Out of the four options available, i.e., 
h, c, e and H$_o$, Laughlin selected $e$} so that, the effective 
charge becomes,
\beg
e_{eff} = e/m
\eeq
where h, c and H$_o$ are kept constant and $m$ is an odd integer.
Therefore, {\bf Laughlin's choice of variables is not unique.}
Let us ask, if some one had selected the velocity of light, out of 
the four variables, then what would have happened? The answer to 
this question is that we would have obtained the particles faster 
than light.

\noindent{\bf3. ~~ Exactness}

     The projection of $\psi_m$ for three and four particles 
onto the lowest energy eigen state of angular momentum 3$m$ for 
m=3 and m=5 states is about 0.98 which shows the exactness of 
the wave functions. Therefore, the exactness of the calculation 
is not affected by changing the variable from $e$ to $H$, h or c.
The calculation is exact for any of the four variables. The total 
energy per particle can be written in terms of the distribution 
function of the one-component plasma (OCP),
\beg
U_{tot}=\pi \int_o^\infty{e^2\over r}
[g(r)-1]rdr\simeq (4/3\pi-1)2e^2/R
\eeq
where integration domain is a disk of radius 
R=($\pi\sigma_m$)$^{-1/2}$.
At $\Gamma$=2, g(r) = 1- exp[-(r/R)$^2$] which shows that for
 $m<$9, the U$_{tot}$ is deeper than for charge density waves. 
The excitations of $\psi_m$ are created by piercing the fluid at 
$z_o$ with an infinitely thin solenoid and passing through it a 
flux quantum $\Delta\varphi$=hc/e, adiabatically. This treatment 
does not treat the magnetic field correctly. Here again,
\beg
field\times area =hc/e
\eeq
so that the variable is e$\times$field$\times$area and hence 
the entire blame can not be thrown on $e$. It may appear that 
changing the $e$ will automatically take care of the electric 
vector {\bf E} and the magnetic field {\bf H} of the 
electromagnetic field but there is {\bf H} in the flux 
quantization which is unchanged. A change in the charge density 
should change both {\bf E} and {\bf H}. According to the adiabatic approximation, the flux quantum produces the quasiparticles. If 
this approximation is not valid, field will be produced not the quasiparticles. In fact the variables, $h$ and $c$ are also very 
good candidates. Increased value of the Planck's constant will 
increase the cyclotron absorption energy and increased $c$ will 
give tycheons: noncausal faster than light particles. The effect of passing a flux quantum is to change the single-body wave function
from,
\beg
(z-z_o)^mexp(-{1\over4}|z|^2)\,\,\,\,to\,\,\,\,(z-z_o)^{m+1}exp(-{1\over4}|z|^2).
\eeq
An approximate representation of these states is chosen by Laughlin 
for the quasielectron and the quasihole. Laughlin writes 
$|\psi^{+z_o}|^2$ as exp(-$\beta\phi\prime$) with $\beta$=1/$m$ 
and $\phi\prime$=$\phi$-2$\Sigma_lln|z_l-z_o|$ where $\phi\prime$ describes an OCP interacting with a phantom point charge at $z_o$. 
The plasma screens this phantom by accumulating an equal and 
opposite charge near $z_o$. The particle charge in the plasma 
is 1, rather than $m$ so the accumulated charge is 1/m, i.e., 
me$_{eff}$ = 1. The energy required to create a particle of Debye 
length $a_o/\sqrt{2}$ is given by,
\beg
\Delta_D= {\pi\over4\sqrt{2}}{e^2\over m^2a_o}.
\eeq

Note that the quantity which enters is again $m^2a_o$ and not just 
$e/m$ so that the charge can be changed from $e$ to $e/m$ or the 
charge may be kept constant at $e$ and $a_o$ changed to $m^2a_o$. Therefore, two equivalent possibilities exist. One is to define the effective charge, e$_{eff}$=$e/m$ and the other is to keep $e$ 
unchanged and change $a_o$ to $a_{eff}$= $m^2a_o$. The state 
described by $\psi_m$ is incompressible because compressing it 
is equal to injecting particles and particles carry charge so the fractional charge will be destroyed. The incompressibility is 
not built in the hamiltonian and is extranuously imposed. It is 
easy to impose an external boundary condition that there is incompressibility but in the expression (14) such an 
incompressibility can not be enforced. Therefore, the 
incompressibility enforcement is not contained in the theory. 
The compressibility creates sound waves so the incompressibility 
causes the sound to be absent. This is equivalent to eliminating phonons, which has not been done, so the resistivity can touch zero value. In the BCS theory of superconductivity, the zero resistance 
is obtained not by introducing incompressibility but by eliminating phonons which makes the electrons attactive. We do not discuss the impurities or rotations because these are not contained in our hamiltonian. The Hall conductance is (1/m)e$^2$/h so that the 
effective charge is e$_{eff}$=(1/m)e and the factor e/h is caused 
by the units. The origin of this effective charge is not clear 
because of the other candidates and for m=3, the charge e/3 is not generated by the hamiltonian discussed by Laughlin but the 
antisymmetry surely requires that m is odd, such as 1, 3 , 5, 7, ... etc. Therefore, odd aspect is a result of antisymmetry. Laughlin
does not use spin so that there is spin-charge decoupling 
automatically built in the calculation. The ${\bf spinless}$ 
electrons may give unphysical results. Below eq.(13) it is argued 
that introducing a flux quantum $\Delta\phi$=hc/e results into accumulation of charge 1/$m$. In fact, the charge need not 
accumulate and only the field is modified as in (8). Therefore, 
the factor 1/$m$ does not determine the charge of the excitations. 
In a different context Laughlin has pointed out that the bulk 
modulus of the Hartree-Fock ground state was ${\it zero}$. The bulk
modulus is actually finite, as is usually the case for Jastrow-type trial wave functions for helium as admitted in an errata$^2$.

\noindent{\bf4.~~ Missing charge}.

   If we agree to the choice of the charge and change it from 
$e$ to $e$/3, then what happened to the remaining charge of 2/3?
If we say that $e$ remains $e$ and $H$ changes to $H$/3, then we 
need not look for the missing charge. The wave function (4) does 
not conserve charge. Laughlin does not provide any prescription
to find the missing charge.

\noindent{\bf 5.~~Conclusions}

     The exact wave function, the excitations of which are 
fractionally charged does not determine the quasiparticle charge uniquely. The blame of the fraction need not be thrown on the quasiparticle charge. The introduction of the alternative choice of variable, instead of charge, does not destroy the exactness of the calculation. The magnetic field of the flux quantization condition 
has been left out.

     It is clear that Laughlin's wave function can not explain the experimental observations of the quantum Hall effect. It may be 
pointed out that the composite fermion model (CF) which requires 
that the flux quanta be attached to the electron is also not correct.$^{3-5}$

     In the expression (13) of Arovas et al$^6$ the quantity $e^*A_{\phi}$ can be replaced by (e/m)A$_{\phi}$. Now, multiply
$A_{\phi}$ by $g$  so that the quantity of interest becomes (e/m)gA$_{\phi}$ with g=1. It is clear that (e/m) and (g/m) are
not resolvable because (e/m)gA$_{\phi}$ is exactly equal to
e(g/m)A$_{\phi}$. Therefore, whether the charge should be 
fractionalized or the unit flux should be corrected are not 
resolved. If charge is corrected, then fractional statistics 
will come otherwise the unit flux will be changed. This means 
that the fermion need not obey the ``fractional statistics" and 
$\phi_o$ will be modified.

     The correct theory of the quantum Hall effect is given in ref.7.
\newpage

\noindent{\bf6.~~References}
\begin{enumerate}
\item R. B. Laughlin, Phys. Rev. Lett. {\bf50}, 1395(1983).
\item R. B. Laughlin, Phys. Rev. Lett. {\bf60}, 2677 (1988);{\bf61}, 379(E)(1988).
\item M. I. Dyakonov, cond-mat/0209206.
\item B. Farid, cond-mat/0003064.
\item K. N. Shrivastava, cond-mat/0209666.
\item D. Arovas, J. R. Schrieffer and F. Wilczek, Phys. Rev. Lett. {\bf53}, 722 (1984).
\item K. N. Shrivastava, Introduction to quantum Hall effect,\\ 
      Nova Science Pub. Inc., N. Y. (2002).
\end{enumerate}
\vskip0.1cm
Note: Ref.7 is available from:\\
 Nova Science Publishers, Inc.,\\
400 Oser Avenue, Suite 1600,\\
 Hauppauge, N. Y.. 11788-3619,\\
Tel.(631)-231-7269, Fax: (631)-231-8175,\\
 ISBN 1-59033-419-1 US$\$69$.\\
E-mail: novascience@Earthlink.net

\vskip0.5cm

\end{document}